\documentstyle[epsfig,a4wide]{article}
\large\normalsize 

\begin{document}

\begin{center}

\title{\textbf{SQUEEZED STATES IN BLACK-HOLE EVAPORATION BY ANALYTIC CONTINUATION}}

\author{\textbf{Andrew N. S. J. Farley} \\
\\
DAMTP, Centre for Mathematical Sciences, University of Cambridge,\\
Wilberforce Road, Cambridge CB3 0WA, UK.\\
\\
A.N.S.J.Farley@damtp.cam.ac.uk }

\maketitle

\end{center}

\large

\vspace{1cm}

\begin{abstract}

\large

We compute the semi-classical quantum amplitude to go from an initial
spherically symmetric bosonic matter and gravitational field
configuration to a final radiation configuration, corresponding to the
relic Hawking radiation from a non-rotating, chargeless black hole which
evaporates completely. This is obtained via the classical action
integral which is solely a boundary term. On discretising the
classical action, the quantum amplitude can be expressed in terms of
generalised coherent states of the harmonic oscillator. A
squeezed-state representation is obtained by complexifying the proper
time separation T at spatial infinity between the initial and final
space-like hypersurfaces. Such a procedure is deemed necessary as the two-surface problem for Dirichlet boundary data and wave-like
perturbations is not well posed. We find that infinitesimal rotation
into the lower complex T plane is equivalent to a highly-squeezed
final state for the relic radiation, similar to the relic
gravitational-wave background in cosmology. This final state is a pure
state, and so the unpredictability associated with the final momentarily-naked singularity is avoided. The cosmological analogy is the
tunnelling from an initial smooth Euclidean or timeless state to a classical
universe. The high-squeezing limit corresponds to a final state of the
Hawking flux which is indistinguishable from a stochastic collection of
standing waves. The phases conjugate to the field amplitudes are
squeezed to discrete values. We also discuss the entropy of the final radiation in the high-squeezing limit.

\end{abstract}

\newpage

\section{Introduction}

Much has been achieved in the application of Feynman's
quantum-amplitude formalism to black-hole evaporation. In a
semi-classical approximation, Hartle
and Hawking~\cite{hartle} pioneered this technique, proving, by analytic
continuation in the complexified space-time, that the
amplitude for a black hole to emit a particle was proportional to the
amplitude for the black hole to absorb a particle. The proportionality
factor is just a Boltzmann factor, whence one determines that a
black hole of initial mass $M_{I}$ can come into equilibrium with the exterior
radiation only at the Hawking temperature $T_{H}=\hbar c^{3}/8\pi
GM_{I}k_{B}.$ Gibbons and Hawking~\cite{gibbhaw} again employed
semi-classical path-integral methods to re-derive Hawking radiance and the Bekenstein-Hawking entropy formula. 

In this paper, we determine the semi-classical wave functional of
specified boundary data in the context of classical-like coherent
states and purely quantum-mechanical squeezed states. The emphasis will be on the final radiation which remains
after a black hole has evaporated completely, and its comparisons with
the relic Cosmic Microwave Background Radiation (CMBR) induced by cosmological
perturbations. 

It is well-known that particle creation by black holes has many similarities with cosmological particle creation, despite the lack of asymptotic
 flatness in the cosmological scenario. Cosmological and black-hole particle creation both require a time
dependence in the metric. The time-dependent background gravitational
 field (also known as the `pump' field) is a result of the non-linearity of
 Einstein's equations, which is displayed through the coupling of the
 gravitational wave field, say, with the evolving background
 space-time. One is also aware of cosmological No-Hair Theorems,
 analogous to the black-hole case, as well the presence of
 singularities, at some finite time in the past in the cosmological
 case, and hidden behind event horizons in the black-hole case.

Rather than employ a time-dependent background metric, black-hole evaporation can be interpreted as a tunnelling process,
whereby the strong gravitational field in the neighbourhood of the
future event horizon promotes vacuum fluctuations into real
particles. Positive-energy particles are detected by observers at
infinity, while negative-energy particles tunnel through the event horizon
reducing the mass of the black hole. Eventually the black hole
presumably disappears completely (we assume in this work that there
are no black-hole relics) leaving only radiation and an almost flat
space-time. The singularity inside the black hole is momentarily
naked prior to total disappearance~\cite{HAW4}, and is
deemed to take with it the information about possible collaspe
configurations which created the black hole. This is the
information-loss paradox.  

In cosmology, near the start of inflation, the quantum vacuum state of
each particle with oppositely directed momenta and short wavelengths is the adiabatic ground state. This is related to the assumption that the
universe was in a maximally-symmetric state at some time in the
past which does not contain a curvature singularity~\cite{HH1}. Due to the accelerated expansion of the universe
during inflation, quantum fluctuations are amplified into
macroscopical or classical perturbations. The early-time fluctuations
lead to the formation of large-scale structure in the universe, and
also contribute to the temperature anisotropies in the CMBR. For modes
whose wavelength is much greater than the Hubble radius, the final state for
the perturbations is a highly-correlated two-mode squeezed
state, with pairs of field quanta produced at late
times with opposite momenta~\cite{grishchuk}. Tensor fluctuations in
the metric, for example, are predicted to give rise to relic gravitational waves. Electromagnetic waves cannot be squeezed during cosmological expansion
in the same way the tensor (metric) perturbations are, because they do not
interact with the external gravitational field in the same way as gravitational waves do.

For the non-rotating, chargeless black hole we shall be
considering, the only parameter characterising the black hole is its
initial mass $M_{I}$. In the adiabatic approximation for evaporating
black holes, for much of the evaporation, frequencies typically exceed $\frac{|\dot{m}|}{m}$, the inverse time-variation
scale for the black-hole mass. That is, the wave period is much smaller
than the timescale of variations of the background gravitational
field. The black hole interacts negligibly with the emitted particles,
and the time between successive emissions is comparable with the black
hole mass~\cite{page}. Prior to the final disappearance of the
hole, however, particle frequencies are of the order $m$, and the
frequency of variations of the background space-time is comparable to
particle frequencies. The particles then interact strongly with the
evanescent black hole. 

In the cosmological case, a natural length scale is the time-dependent
inverse Hubble parameter $H^{-1}=\frac{a}{\dot{a}}$ ($a(t)$ is the
Friedmann-Robertson-Walker scale factor), with the adiabatic
approximation $k\gg H$, or such that the wavelengths of the
perturbations are less than the Hubble radius $aH^{-1}$. When the
wavelength is comparable with or longer than the Hubble radius, the
amplification of the zero-point fluctuations takes place. In addition, both
the ADM mass and Hubble parameter control the redshifting of the
radiation in the background space-time.

Due to the Schr\"{o}dinger evolution, cosmological perturbations (rotational, density and gravitational) in
an initial vacuum state are transformed into a highly-squeezed vacuum
state with many particles having a large variance in their amplitude
(particle number) and small (squeezed) phase variations. At small wavelengths, the squeezing of cosmological
perturbations may be suppressed, while it should be present at long
wavelengths, particularly for gravitational
waves~\cite{Grishchuk:1995ia}. These perturbations also induce the
large-angular-scale anisotropies observed in the CMBR. Their
wavelengths today are of the order of or greater than the Hubble
radius. The amplification of the initial zero-point fluctuations gives
rise to standing waves with a fixed phase, rather than travelling
waves. The relic perturbations in the high-squeezing or WKB limit can
be described as a stochastic collection of standing waves. However, it has been
suggested that the squeezed-state formalism engenders no new
physics~\cite{Albrecht:1992kf}. 

A prominent theme of this paper is the squeezed-state
formalism applied to black-hole evaporation, with reference to the
comparable inflationary-cosmology scenario. Grishchuk and
Sidorov~\cite{grishchuk} first achieved a squeezed-state
representation for Hawking radiation. Their approach, however,
referred to final quantum states describing the particles escaping to
infinity and those falling through the event horizon. In this paper,
the squeezed-state approach is related to the radiation which remains
after a black hole has evaporated completely. In an adiabatic approximation,
the fixed phases correspond to discrete frequencies in the remnant
Hawking radiation from the evanescent black hole. This high-squeezing
feature, we argue, can potentially be observed. The standing
waves of the highly-squeezed final state of the Hawking radiation
originates from choosing to set Dirichlet boundary data on initial
and final space-like hypersurfaces. Such a problem is not well
posed, and so we propose complexifying the proper time interval at
spatial infinity in order to obtain a well-posed problem. This, we
believe, has the effect of avoiding the problems associated with the
final naked singularity which may be present prior to the final
disappearance of the black hole. Thus, we conclude that a
highly-squeezed final state, which is a pure state, is related to the
avoidance of the naked singularity, and its associated unpredictability.

After a brief summary of our two-surface method in
Section \ref{twosurface}, in Section \ref{amplitude} we write the
quantum amplitude for linearised spin-$0,1,2$ fields in a unified
form. Coherent states and generalised coherent states are related to
the quantum amplitude in Section \ref{coherent}. Section \ref{squeeze}
introduces the squeezed-state formalism into the theory {\it via} a
complexification technique. In Section \ref{classical}, we discuss issues
of entropy and classical behaviour. Throughout we employ Planckian
units $G=c=\hbar =1.$

\section{Two-Surface Formulation}\label{twosurface} 

This section is a summary of results which will be elaborated on in ref.~\cite{me}.

Our approach to black-hole evaporation is through a two-surface
 boundary-value problem. We consider the quantum-mechanical decay of
 a chargeless, non-rotating black hole into almost flat space-time and
 purely outgoing radiation. Data for
 spin-$0,\frac{1}{2},1,\frac{3}{2},2$ perturbations is set on an
 initial space-like hypersurface $\Sigma_{I}$, just prior to the
 collapse of an initially diffuse (spherical) `star' to a black hole,
 and a final space-like hypersurface $\Sigma_{F}$, sufficiently far to
 the future of $\Sigma_{I}$ that it catches the relic Hawking
 radiation. The black hole evaporates completely into predominantly
 massless particles, with the final total (ADM) energy equal to the
 initial mass $M_{I}$.

The classical action $S^{(cl)}$, i.e., the action evaluated at the
solution of the field equations, reduces to solely a boundary term,
where the boundary comprises $\Sigma_{I}$, $\Sigma_{F}$ and a
time-like boundary $\Sigma^{\infty}$ ($\{r=R_{\infty}\}$) located at
large radius connecting $\Sigma_{I}$ to $\Sigma_{F}$, the proper time
separation at spatial infinity being denoted by $T$. For simplicity, we considered initial data of very weak perturbations, i.e., such that the matter
and gravitational fields were initially spherically symmetric. The
final data comprises non-zero perturbations corresponding to the relic
radiation after the black hole has disappeared. 

Let us specialise to the integer-spin perturbations; the fermions will be
considered in a future paper. For the wave-like field equations
satisfied by the perturbations, the `Dirichlet' boundary-value problem
on a purely Lorentzian background space-time is not well-posed~\cite{garabedian}. That
is, there may be no solution to the field equations with this boundary
data, or an infinite number of solutions. Adopting a `Euclidean'
strategy, we rotate $T$ into the lower complex plane. Consequently,
one might then expect there to be a unique classical solution to the
nearly-Lorentzian field equations. This is just Feynman's
$\pm i\varepsilon$ prescription in quantum field theory.

In a neighbourhood of $\Sigma_{I}$ and $\Sigma_{F}$, one may employ
an adiabatic approximation, whereby most of the radiation frequencies
exceed the rate of change of the background space-time. One may then `Fourier' decompose the perturbations into
standing waves, so that the boundary conditions on $\Sigma_{I}$ and
$\Sigma_{F}$ are satisfied. For example, the real linearised massless
scalar field $A^{(1)}(x)$ which vanishes on $\Sigma_{I}$ and is
non-zero on $\Sigma_{F}$ is written as
\begin{equation}
A^{(1)}(x)=\frac{1}{r}\sum_{l=0}^{\infty}\sum_{m=-l}^{l}\int_{-\infty}^{\infty}dk\,a_{0klm+}\xi_{0kl+}(t,r)\frac{\sin(kt)}{\sin(kT)}Y_{lm}(\Omega).\label{q1'}
\end{equation}
The background line-element is spherically-symmetric on a time
average and has the time-dependent Schwarzschild-like form
\begin{equation}
ds^{2}=-e^{b(t,r)}dt^{2}+e^{a(t,r)}dr^{2}+r^{2}d\Omega^{2},\label{q5}
\end{equation}
where $d\Omega^{2}$ is the two-sphere line-element. In the limit of
wavelengths small compared to the background radius of curvature, the
background geometry at late times is approximately a Vaidya space-time. The
$\{Y_{lm}(\Omega)\}$ are spherical harmonics with angular momentum $l$
and azimuthal angular momentum $m$. The $\{\xi_{0kl+}(t,r)\}$ are
`radial' functions which are functions of $r$ and slowly-varying
functions of $t$ in an adiabatic approximation. The $\{a_{0klm+}\}$ are a set of time-independent
amplitudes (see below). Only one set of amplitudes is associated with
the final state of the radiation as we suppose that there is no event
horizon since the black hole disappears completely. A decomposition
similar to eqn.(\ref{q1'}) for the spin-1 and spin-2
perturbations is also possible -- see ref.~\cite{me}. The boundary-value problem for the
fermionic fields differs from the bosonic case due to the first-order nature of their actions.

By taking the limit $R_{\infty}\to\infty$,
one can normalise the radial functions on $\Sigma_{F}$, assuming
regularity at the origin $\{r=0\}$, which implies that the radial
functions are real, and a linear combination of plane waves at
infinity. The classical action can then be written as a single
integral over frequency. For the spin-$0, 1, 2$ perturbations, the
classical action has an infinite number of discrete simple poles along
the real frequency axis at the frequencies
$\omega_{n}=\frac{n\pi}{|T|}, n=1,2,...$ This is a manifestation of
the ill-posedness of the two-surface boundary-value problem. Through our `Euclidean'
strategy, we can avoid the poles along the real frequency axis and
subsequently obtain a probabilistic interpretation for the final
boundary data. This is because the classical action acquires real and
imaginary parts in the complexification procedure. At the end of our
calculations, we take the limit that $T$ approaches the real axis, a
procedure intimately connected with the squeezed-state formalism. In the next section, we will compute the quantum
amplitude for the integer-spin fields as a stepping stone to computing the probability density.

Additional papers~\cite{me} matched our two-surface approach with the method of
Bogoliubov coefficients tailored to describing the radiation incoming
from and outgoing to the null hypersurfaces ${\mathcal{I}^{-}}$ and
${\mathcal{I}^{+}}$ respectively.

\section{The Quantum Amplitude}\label{amplitude}

\noindent

For the massless spin $s=0, 1, 2$ perturbations, we find that the
quantum amplitude or wave functional $\Psi[\{a_{sklmP}\};T]$ is given by~\cite{me}
\begin{equation}
\Psi[\{a_{sklmP}\};T] = N\exp\left(iS^{(cl)}[\{a_{sklmP}\};T]\right),\label{q1}
\end{equation}
where $\{a_{sklmP}\}$ denotes a set of `Fourier'-like amplitudes for
perturbations with spin $s$, frequency $k$, angular momentum $l$,
azimuthal angular momentum $m$ and polarisation $P$, and $S^{(cl)}$ is
the linearised total classical action. (For a massless scalar field,
the polarisation can refer to the real and imaginary parts of a single
complex massless scalar field.) Further, $N$ is a $T$-dependent
prefactor given below. We emphasise that eqn.(\ref{q1}) is just a
semi-classical approximation to the full quantum amplitude, which is
given by a path integral over field configurations which match the
specified boundary data. For the simplest initial Dirichlet boundary
condition that the initial perturbations are very weak, we found that 
\begin{eqnarray}
S^{(cl)}[\{a_{sklmP}\};T]\!\! & = & \!\!\frac{1}{4\pi}\!\sum_{s}\sum_{l=s}^{\infty}\sum_{m=-l}^{l}\sum_{P=\pm}w_{s}\frac{(l-s)!}{(l+s)!}P\!\int_{0}^{R_{\infty}}\!dre^{\frac{1}{2}(a-b)}\xi_{slmP}\partial_{t}\xi^{*}_{slmP}\!\!\mid_{\Sigma_{F}}\nonumber\\
& - & \frac{1}{2}M_{I}T\nonumber\\
& = & \!\!\!\sum_{slmP}(-1)^{s}w_{s}\frac{(l-s)!}{(l+s)!}\int_{0}^{\infty}\!dk\,k
|z_{sklP}|^{2}|a_{sklmP} +(-1)^{s}Pa_{s,-klmP}|^{2}\cot(kT)\nonumber\\
& - & \frac{1}{2}M_{I}T.\label{q2}
\end{eqnarray}
 The final term in eqn.(\ref{q2}) comes from the time-like boundary
$\Sigma^{\infty}$ and gives a plane-wave representation for black
holes of fixed ADM mass $M_{I}$ in the absence of perturbations. In
eqn.(\ref{q2}), $w_{0}=2\pi, w_{1}=\frac{1}{4}, w_{2}=\frac{1}{8}$,
and $a_{sklmP}\!=\!P(-1)^{m}a_{s,-kl,-mP}^{*}.$ 

The functions $\{\xi_{slmP}(t,r)\}$ satisfy in an adiabatic approximation $k\gg \frac{1}{2}|\dot{a}-\dot{b}|$~\cite{me}
\begin{equation}
e^{\frac{1}{2}(b-a)}\partial_{r}(e^{\frac{1}{2}(b-a)}\partial_{r}\xi_{slmP})-\partial_{t}^{2}\xi_{slmP}+V_{slP}(t,r)\xi_{slmP}=0,\label{q3}
\end{equation}
where $V_{slP}(t,r)$ is a spin-dependent potential:
\begin{equation}
V_{slP}(t,r)=\frac{e^{b(t,r)}}{r^{2}}\left[l(l+1)+(1-s^{2})\frac{2m(t,r)}{r}\right],
\end{equation}
for $s=0,1$ and odd-parity spin-2 perturbations, and a more
complicated term for the even-parity metric perturbations which will
not be needed in this paper. The effect of the
back-reaction on the metric in the adiabatic approximation is just to
replace the Schwarzschild constant mass with the time- and
radially-dependent mass $m(t,r)$ in $V_{slP}$ where $e^{-a(t,r)}=1-\frac{2m(t,r)}{r}$. The spin-0 and spin-2 problems are very similar. In cosmology,
primordial density perturbations and the relic gravitational waves are
also described by similar equations.  The coefficients $\{z_{sklP}\}$
relate to the boundary conditions at spatial infinity and the
regularity conditions at the origin for the radial part of the real functions
$\{\xi_{slmP}(t,r)\}$~\cite{me}. In the first term of eqn.(\ref{q2}), one must
understand that the limit $R_{\infty}\to\infty$ is implied.

 Equation (\ref{q1}) can be interpreted as giving a
`coordinate'-representation amplitude for a set of final field
configurations $\{a_{sklmP}\}$ labelled by `quantum' numbers $\{sklmP\}$, given that on the hypersurface $\Sigma_{I}$ $\{t=0\}$ the perturbations vanished,
i.e., the metric and background matter were spherically
symmetric. Hence $|\Psi[\{a_{sklmP}\};T]|^{2}$ is a (conditional) probability
density of finding the field in a set of final configurations
$\{a_{sklmP}\}$ at asymptotic proper time $T$. 

To proceed further, let us discretise the frequency integral in
eqn.(\ref{q2}), with $\{k_{j}>0\}$ the eigenfrequencies of the final
radiation located in a spatial volume on $\Sigma_{F}$ --- the spatial
momenta. An explicit expression for the $\{k_{j}\}$ will not be
needed, although the continuum limit is recovered in the limit $R_{\infty}\to\infty$. Denoting $j=\{slmP\}$, we find that eqn.(\ref{q1}) can be
written as (see also Appendix A)
\begin{eqnarray}
\Psi[\{A_{j}\};T] & = & \hat{N}e^{-i\frac{1}{2}M_{I}T}\prod_{j}\frac{1}{2
i\sin(k_{j}T)}\exp\left(\frac{i}{2}\Delta k_{j} k_{j}|A_{j}|^{2}\cot(k_{j}T)\right)\nonumber\\
 & = & \hat{N}e^{-i\frac{1}{2}M_{I}T}e^{-\frac{1}{2}\sum_{j}\Delta k_{j} k_{j}|A_{j}|^{2}}\prod_{j}\sum_{n=0}^{\infty}e^{-2iE_{n}T}L_{n}(k_{j}\Delta k_{j}|A_{j}|^{2}),\label{q6}
\end{eqnarray}
 where
\begin{equation}
|A_{j}|^{2}=2(-1)^{s}w_{s}\frac{(l-s)!}{(l+s)!}|z_{j}|^{2}|a_{j}+(-1)^{s}Pa_{s,-k_{j} lmP}|^{2},\label{q7}
\end{equation}
and $E_{n}=k_{j}(n+\frac{1}{2})$ is the quantum energy of the linear
harmonic oscillator. Note the dependence of the quantum amplitude on
$|A_{j}|$; that is, it is `spherically symmetric'. The $\{L_{n}\}$ are Laguerre
polynomials with
\begin{equation}
L_{n}^{(m-n)}(x)=\sum_{p=0}^{n}\left( \begin{array}{c}
m \\
n-p \\ \end{array}\right)\frac{(-x)^{p}}{p!}\label{q8}
\end{equation}
the associated Laguerre polynomials, and $L_{n}^{(0)}(x)=L_{n}(x)$~\cite{grad}. The $\{L_{n}(x)\}$ satisfy the completeness relation
\begin{equation}
\sum_{n=0}^{\infty}e^{-\frac{1}{2}x}L_{n}(x)e^{-\frac{1}{2}y}L_{n}(y)=\delta(x-y).\label{q9}
\end{equation}
The Laguerre polynomials $\{L_{n}(|z|^{2})\}$, where $z=x+iy$, cannot
be written as a product of two decoupled wave functions of $x$ and $y$
in an excited state (due to pair correlations~\cite{Gasperini:1993mq} in a
quantum interpretation), but we can write~\cite{erdelyi}
\begin{equation}
L_{n}(x^{2}+y^{2})=\frac{(-1)^{n}}{2^{2n}n!}\sum_{p=0}^{n}\left( \begin{array}{c}
n \\
p \\ \end{array}\right)H_{2p}(x)H_{2n-2p}(y).\label{q10}
\end{equation}
The wave functional $\Psi[\{A_{j}\};T]$ is proportional to a
product of delta functions when $k_{j}T=p\pi$ ($p$ being an integer),
following from the completeness relation eqn.(\ref{q9}) -- see
Appendix A. At these
focal points, the density of paths -- effectively the prefactor in the
first line of eqn.(\ref{q6}) -- diverges. 

\section{Coherent States}\label{coherent}

The Schr\"{o}dinger picture wave functions 
\begin{equation}
\Psi_{nj}(x_{j},T)=\frac{N}{\pi}e^{-\frac{1}{2}x_{j}}e^{-2iE_{n}T}L_{n}(x_{j}),\label{q11}
\end{equation}
where $x_{j}=k_{j}\Delta k_{j}|A_{j}|^{2},$ have a strong connection
with the exact solution of the forced-harmonic-oscillator problem~\cite{schwinger}. In this theory, one considers a 1-d harmonic
oscillator with mass $\mu$ and frequency $\omega,$ which is acted on
by an external force $F(t).$ In this case, the Hamiltonian has the
form
\begin{equation}
H=\frac{1}{2}(p^{2}+q^{2})+qF(t).\label{q12}
\end{equation}
Suppose that for $t_{0}<t<T$ this force
is non-vanishing so that asymptotic states are free oscillator
states. We want to calculate the probability amplitude  $A_{km}$ to
make a transition from the free oscillator state $|m\!>$ (with $m$
particles) at time $t'<t_{0},$ before the force begins to act, to the
free oscillator state $|k\!>$ at time $t>T,$ after the force has ceased. Set 
\begin{eqnarray}
z & = & \frac{|\beta|^{2}}{2\mu\omega},\label{q9'} \\
\beta & = & \int_{t_{0}}^{T}dt\,F(t)e^{-i\omega t},\label{q10}
\end{eqnarray}
effectively the `Fourier' transform of the force. It has been shown
that~\cite{roy}~--~\cite{glauber1}\\ ($m\geq k$)
\begin{equation}
A_{km}=e^{i\lambda}e^{-\frac{1}{2}z}\left (\frac{k!}{m!}\right
)^{\frac{1}{2}}\left (\frac{i\beta}{\sqrt{2\mu\omega}}\right
)^{m-k}L_{k}^{(m-k)}(z)\label{q11'},
\end{equation}
where  $\lambda$ is a real phase and $A_{km}$ is symmetric in $k$ and
$m$. An adiabatic approximation $(z\ll 1)$ indicates that a state
which begins as $|m\!>$ must end up in the same state at late times after the
time-dependent force has been removed. Then
\begin{equation}
A_{kk}=e^{i\lambda}e^{-\frac{1}{2}z}L_{k}(z).\label{q13}
\end{equation}
The probability that there be no change in the number of particles is
therefore $|A_{kk}|^{2}=e^{-z}(L_{k}(z))^{2}$. Apart from the introduction of mode labels $\{j\}$ and a
necessary reinterpretation for $z,$ these amplitudes are effectively
the semi-classical wave functions Eqn.(\ref{q11}) derived from our boundary-value problem.

A brief derivation of eqn.(\ref{q11'}) in the context of the
coherent-state representation~\cite{glauber1} will be useful for the
calculations in Section \ref{GCS}. Coherent states
$|\alpha>$ can be regarded as displaced vacuum states~\cite{glauber2}:
\begin{equation}
|\alpha>=D(\alpha)|0>,\label{q14}
\end{equation}
where $D(\alpha)=e^{\alpha a^{\dagger}-\alpha^{*}a}$ is a unitary
displacement operator ($D^{\dagger}(\alpha)=D^{-1}(\alpha)=D(-\alpha)$)
and the states $|\alpha>$ are eigenstates of the annihilation operator
$a$ with complex eigenvalue $\alpha$. They are the closest states to
classical states in that they attain the minimum demanded by the
uncertainty principle. Thus, coherent states may also be useful when
one has a large number of particles, as in the classical limit, and
when one has some phase information about the state. Coherent states
form an over-complete set, but are not orthogonal. In terms of Fock
number states $|n>=\frac{(a^{\dagger})^{n}}{\sqrt{n!}}|0>$~\cite{glauber1},
\begin{equation}
|\alpha>=e^{-\frac{1}{2}|\alpha |^{2}}\sum_{n=0}^{\infty}\frac{\alpha^{n}}{\sqrt{n!}}|n>.\label{q15}
\end{equation}
The coherent state corresponding to $\alpha =0$ is the unique state of
the oscillator, namely, the Fock state $|n>$ with $n=0.$ Thus, if the
system started in a vacuum state, the amplitude to then find it in a
coherent state $|\alpha >$ is
\begin{equation}
<0|\alpha>=<0|D(\alpha)|0>=e^{-\frac{1}{2}|\alpha|^{2}},\label{q15'}
\end{equation}
up to a phase.

From the properties of displacement operators~\cite{glauber2}, then
\begin{eqnarray}
D(\xi)|\alpha> & = & D(\xi)D(\alpha)|0>\nonumber\\
& = & D(\xi+\alpha)|0>e^{\frac{1}{2}(\xi\alpha^{*}-\xi^{*}\alpha)}\nonumber\\
& = & |\xi+\alpha >e^{\frac{1}{2}(\xi\alpha^{*}-\xi^{*}\alpha)}.\label{q16}
\end{eqnarray}
For later reference, one can show that
\begin{equation}
D^{\dagger}(\alpha)D(\mu)D(\alpha)=D(\mu)e^{\alpha^{*}\mu-\alpha\mu^{*}}.\label{q16'}
\end{equation}
Using eqns.(\ref{q15}) and (\ref{q16}), 
\begin{equation}
<m|D(\xi)|\alpha>=\frac{1}{\sqrt{m!}}(\xi+\alpha)^{m}e^{-\frac{1}{2}(|\alpha |^{2}+|\xi|^{2}+2\xi^{*}\alpha)}.\label{q17}
\end{equation}
Also from eqn.(\ref{q15}),
\begin{equation}
<m|D(\xi)|\alpha>=e^{-\frac{1}{2}|\alpha |^{2}}\sum_{n=0}^{\infty}\frac{\alpha^{n}}{\sqrt{n!}}<m|D(\xi)|n>.\label{q18}
\end{equation} 
Equating eqns.(\ref{q17}) and (\ref{q18}) then gives
\begin{equation}
(1+y)^{m}e^{-y|\xi |^{2}}=e^{\frac{1}{2}|\xi |^{2}}\sum_{n=0}^{\infty}\sqrt{\frac{m!}{n!}}\xi^{n-m}y^{n}<m|D(\xi)|n>.\label{q19}
\end{equation} 
The generating function for the associated Laguerre
polynomials~\cite{grad}
\begin{equation}
(1+y)^{m}e^{-yx}=\sum_{n=0}^{\infty}L_{n}^{(m-n)}(x)y^{n}, \,\,\,|y|<1,\label{q20}
\end{equation} 
then implies that the matrix element between initial and final states can be written as
\begin{equation}
<m|D(\xi)|n>=\left(\frac{n!}{m!}\right)^{\frac{1}{2}}\xi^{m-n}e^{-\frac{1}{2}|\xi |^{2}}L_{n}^{(m-n)}(|\xi |^{2}),\label{q21}
\end{equation} 
which is eqn.({\ref{q11'}) up to an unimportant phase
factor. Interchanging $m$ and $n$ gives the same result.

\subsection{Generalised Coherent States}\label{GCS}

\noindent

There is an interpretation for these amplitudes in terms of
generalised coherent states $|n,\alpha>$ of the harmonic
oscillator~\cite{saty}. Defining
\begin{equation}
|n,\alpha>=e^{-iE_{n}t}D(\alpha(t))|n>,\label{q22}
\end{equation} 
then in the Fock representation
\begin{equation}
|n,\alpha>=\sum_{m=0}^{\infty}<m|D(\alpha(0))|n>|m>e^{-iE_{m}t}.\label{q23}
\end{equation} 
For generalised coherent states (see eqn.(\ref{q27}) below), the ground
state ($n=0$) is a coherent state and not a vacuum state. Generalised coherent states are to
the coherent states what the Fock states $|n>$ are to the vacuum
state, that is, excited coherent states. In addition, one has
\begin{eqnarray}
I & = & \frac{1}{\pi}\int\,d^{2}\alpha |n,\alpha><n,\alpha |,\label{q24}\\
<n,\beta |n,\alpha> & = & L_{n}(|\alpha-\beta|^{2})e^{\beta^{*}\alpha-\frac{1}{2}(|\alpha|^{2}+|\beta|^{2})},\label{q25}\\
<n,\beta |\psi> & = & \frac{e^{-\frac{1}{2}|\beta|^{2}}}{\pi}\int\,d^{2}\alpha L_{n}(|\alpha-\beta|^{2})e^{\beta^{*}\alpha}e^{-\frac{1}{2}|\alpha|^{2}}<n,\alpha |\psi>,\label{q26}
\end{eqnarray}
for an arbitrary state $|\psi>$, where $$\frac{1}{\pi}\int\!d^{2}\alpha
=\int d(Re\alpha)d(Im\alpha),$$ and $I$ is the identity operator. From eqn.(\ref{q25}) with $\beta =0$, then
\begin{equation}
<n,0|n,\alpha>\,\, \equiv\,\, <n|n,\alpha>\,\,=\,\,e^{-\frac{1}{2}|\alpha|^{2}}L_{n}(|\alpha |^{2}),\label{q27}
\end{equation}
again giving eqn.(\ref{q11}) up to a phase and normalisation. The initial state may be seen not as a vacuum state, but as a Fock state, and the final
state as a generalised coherent state with the same $n$.

 An interesting interpretation for the amplitudes eqn.(\ref{q21}) is
that these are the matrix elements for the transition from the state $|k>$ to $|m>$ under
the influence of a gravitational wave~\cite{Holl}, with the force
$F(t)$ proportional to the Riemann curvature tensor component
$R_{xtxt}(t)$:
\begin{equation}
F(t)=\mu c^{2}lR_{xtxt}(t)=-\frac{1}{2}\mu c^{2}l\partial_{t}^{2}h_{xx}^{TT},\label{q27'}
\end{equation}
where $l$ is the distance between two particles each of mass
$\frac{1}{2}\mu$ along the $x$-axis, $h_{xx}^{TT}$ is the transverse traceless
gravitational wave component~\cite{MTW} and $x$ is the change in
separation of the masses. In the context of black-hole evaporation, one can understand the
time-dependent force as being active during the time-dependent phase
of the collapse to a black hole and its subsequent complete
evaporation. The duration of the `force', therefore, is
comparable to the lifetime of the black hole. A
space-time would begin in an (almost) static state in the far past, where the
perturbation modes were effectively free, pass through an intermediate
time-dependent phase, to end up in a static or, rather,
quasi-stationary state in the far future, where again the perturbation
modes are free. In this `sandwich'-space-time picture, particles are
created by the space-time curvature, or non-adiabatic behaviour of the fields in a time-dependent metric. What the above calculations indicate is an explicit
mathematical connection between the theory of forced harmonic
oscillators and certain amplitudes relating to the dynamical evolution
of black holes. 

\section{Squeezed-State Formalism}\label{squeeze}

In this section, we see how by rotating the proper time at infinity $T$ into the complex plane,
and with spherically symmetric initial matter and gravitational
fields, one obtains a quantum-mechanical squeezed-state interpretation
for the final state of the Hawking radiation.

Grishchuk and Sidorov~\cite{grishchuk} were the first to explicitly
formulate particle creation in strong gravitational fields in terms of
squeezed-states, although the formalism does appear in
Parker's original paper on cosmological particle
production~\cite{parker}. In ref.~\cite{grishchuk}, it was shown that relic
gravitons (as well as other perturbations), created from zero-point quantum fluctuations as
the universe evolves, should now be in a strongly squeezed
state. Squeezing is just the quantum process of parametric
amplification. In ref.~\cite{Albrecht:1992kf} however, the claim
was made that there was no new physics in employing the squeezed-state
formalism. 

\subsection{Squeezed States -- Introduction}

A general one-mode squeezed state (or squeezed coherent state) is
defined as~\cite{shoemaker}
\begin{equation}
|\alpha,z>=S_{1}(r,\varphi)D(\alpha)|0>=S_{1}(z)D(\alpha)|0>=S_{1}(z)|\alpha>,\label{q33}
\end{equation} 
where $D(\alpha)$ is the single-mode displacement operator,
$S_{1}(r,\varphi)$ is the unitary squeezing operator
($S_{1}^{\dagger}(z)S_{1}(z)=S_{1}(z)S_{1}^{\dagger}(z)=1)$ for $|\alpha,z>$:
\begin{equation}
S_{1}(r,\varphi)\equiv S_{1}(z) = e^{\frac{1}{2}(za^{2}-z^{*}a^{\dagger 2})},\label{q35}
\end{equation}
where $z=re^{-2i\varphi}.$ Further,
$S_{1}(r,\varphi)S_{1}(r',\varphi)=S_{1}(r+r',\varphi)$. With
$a|\alpha >=\alpha |\alpha >$, then setting $b=S_{1}aS_{1}^{\dagger}$
implies that $b|\alpha,z>=\alpha |\alpha,z>$ and $b=a\cosh r
+a^{\dagger}e^{2i\varphi}\sinh r.$ The state $|\alpha,z>$ is a Gaussian
packet displaced from the origin in position and momentum space. While
the real squeezing parameter $r\geq 0$ determines
the magnitude of the squeezing, the squeezing angle $\varphi$
($-\frac{\pi}{2}<\varphi\leq\frac{\pi}{2}$), gives
the distribution of the squeezing between conjugate variables.
The operators $a$ and $a^{\dagger}$ are annihilation and creation operators,
respectively. The squeezed vacuum state occurs when $\alpha =0$:
\begin{equation}
|z>\,\equiv |0,z>\,=S_{1}(z)|0>.\label{q36}
\end{equation}
The high-squeezing limit corresponds to $r\gg 1,$ where the state
$|z>$ is highly localised in momentum space. The state with $r=0$ is
the ground state. 

Consider the amplitude $A=<\alpha ,z|D(\mu)|\alpha
,z>=<z|D^{\dagger}(\alpha)D(\mu)D(\alpha)|z>$. One can use
eqn.(\ref{q16'}) to show that $A=<z|D(\mu)|z>e^{2i Im(\alpha^{*}\mu
)}=<0|D(\eta)|0>e^{2i Im(\alpha^{*}\mu
)}$, where $\eta=\mu\cosh r-\mu^{*}e^{2i\varphi}\sinh r,$ and then use $\mu=|\mu |e^{i\phi}$
to show that~\cite{Holl}
\begin{equation}
|A|^{2}=e^{-|\eta |^{2}}=e^{-|\mu |^{2}[\cosh 2r-\cos2(\phi+\varphi)\sinh 2r]},\label{q36'}
\end{equation}
which is independent of the initial displacement $\alpha$. Equation
(\ref{q36'}) will appear when we match our space-like hypersurface
approach with the method of Bogoliubov coefficients~\cite{me}. 

Single-mode squeezed operators do not conserve momentum since they
describe the creation of particle pairs with momentum $k$. Two-mode
squeezed operators, however, describe the creation and annihilation of
two particles (waves) with equal and opposite momenta. A two-mode squeeze operator has the form~\cite{Hu:1993gm} 
\begin{equation}
S_{2}(r,\varphi)=e^{r(e^{-2i\varphi}a_{+}a_{-}-e^{2i\varphi}a_{+}^{\dagger}a^{\dagger}_{-})},\,\,\,S_{2}^{\dagger}=S_{2}^{-1}=S_{2}(r,\varphi
+\frac{\pi}{2}),\label{q36''}
\end{equation}
where $a_{\pm},a_{\pm}^{\dagger}$ are annihilation and creation
operators for the two modes, respectively.

One can also introduce a unitary rotation operator
\begin{equation}
R(\theta)=e^{-i\theta (a^{\dagger}_{+}a_{+}+a_{-}^{\dagger}a_{-})},
\end{equation}
with $0\leq\theta\leq 2\pi$, $R(\theta)R(\theta ')=R(\theta +\theta
'),$ $R|0>=|0>, RaR^{\dagger}=e^{i\theta}a.$ However, rotation does
not influence particle creation~\cite{Hu:1993gm}.

The variances of conjugate operators $\hat{p}$ and $\hat{q}$ are given
by $\Delta\hat{q}=\hat{q}-<\hat{q}>, \Delta\hat{p}=\hat{p}-<\hat{p}>.$
In the squeezing formalism, $\Delta\hat{q}$ and $\Delta\hat{p}$ differ
greatly, while they are equal and the minimum possible for coherent
states, deemed to be the most classical of quantum states. The name `squeezed' refers to the fact that the variance of one variable
in a conjugate pair can go below the minimum allowed by the
uncertainty principle (the squeezed variable), while the variance of
the conjugate variable can exceed the minimum value allowed (the
superfluctuant variable)~\cite{Gasperini:1993mq}~\cite{Gasperini:1992xv}~\cite{Gasperini:1995yd}. The superfluctuant variable is amplified by
the squeezing process, and so it is possible to observe
macroscopically, while the subfluctuant variable is squeezed and
becomes unobservable. In particle production, for example, the
number operator is a superfluctuant variable, while the phase is
squeezed. Squeezed states are essentially purely quantum-mechanical in origin.

Hawking radiation in the squeezed-state representation
was first discussed in ref.~\cite{grishchuk}. The squeeze parameter $r_{j}$ was
related to frequency and the black-hole mass through $\tanh r_{j}
=e^{-4\pi\omega_{j}M}, \theta_{j}=\varphi_{j}$ (see also
refs.~\cite{suresh} and~\cite{Kiefer:2001wn}). The vacuum quantum state in
a black-hole space-time for each mode is a two-mode squeezed
vacuum.

\subsection{Analytic Continuation and the Large-Squeezing Limit}

Consider the Schr\"{o}dinger picture quantum state eqn.(\ref{q6})
associated with the final hypersurface $\Sigma_{F}$, and define
\begin{eqnarray}
\Phi[\{A_{j}\};T] & = & \prod_{j}2i\sin (k_{j}T)\Psi[\{A_{j}\};T]=\hat{N}e^{-i\frac{1}{2}M_{I}T}\prod_{j}\exp\left[\frac{i}{2}\Delta k_{j} k_{j}|A_{j}|^{2}\cot(k_{j}T)\right]\nonumber\\
 & \equiv & \hat{N}\exp\left(iS^{(cl)}[\{A_{j}\};T]\right).\label{q37}
\end{eqnarray}
In order for our two-surface boundary-value problem to be well posed,
we argue that $T$ must be a complex quantity. This is because for real $T,$  the `sum' in eqn.(\ref{q37}) diverges
due to the simple poles on the real-frequency axis at
$k_{j}=\sigma_{n}=\frac{n\pi}{T}, n=1,2,...$, assuming that $k_{j}|A_{j}|^{2}$ remains finite and non-zero near $k_{j}=\sigma_{n}.$ At the frequencies $\{\sigma_{n}\}$, there may be an infinite number of classical paths
(solutions) joining the initial and final data, or perhaps none at
all. Should a solution exist, it may not depend smoothly on the
boundary values. These are features of ill-posed boundary-value
problems. Yet, if $T$ is deformed into the complex plane, the poles along the real-$k_{j}$ axis would be displaced into
the complex-$k$ plane. The complex boundary-value problem itself
guarantees that the classical solution is complex analytic (strong
ellipticity). One then has a reasonable expectation that there will
now be a smooth, unique classical solution of the field equations
joining the initial and final data. Essentially, this is the
$i\varepsilon$ prescription of quantum-field theory. In general, therefore, our background metric $\gamma_{\mu\nu}$ (and hence
$g_{\mu\nu}$) is not a real Lorentzian metric but perhaps a strongly-elliptic
metric. This may permit the path integral over the boundary data to be
well-defined. Even for a complexified $T$, one
still has a time, $|T|$, whose `gradient' is everywhere
future-directed and time-like. By considering the dimensionless
quantity $T/2M_{I},$ one also sees that analytic continuation of the complex quantity $T$ is equivalent to adding a small positive imaginary
part to the mass, as is the case in the Damour-Ruffini tunnelling approach to black-hole evaporation~\cite{DR}.

In terms of the late stages of black-hole evolution, we may use
analytic continuation to tackle the crucial question of whether future
null infinity $\mathcal{I}^{+}$ is a Cauchy hypersurface for
space-time. If we model the evaporation process with a classical
Lorentzian space-time metric, it is well-known that a
momentarily-naked singularity is inevitable~\cite{HAW4}. Cosmic
censorship, a postulate of classical general relativity, is seemingly
transcended through the quantum-mechanical evaporation of the black
hole. Consequently, a Bogoliubov transformation between initial and
final modes would be forbidden. There would then be no one-to-one correspondence between pre-collapse
configurations and post-evaporation configurations, no S-matrix and,
therefore, there will be evolution from a pure initial state to a final density
matrix. Our reasoning is that a non-singular asymptotically-flat
initial state on a suitable Cauchy hypersurface can never evolve to a
final (even momentarily) naked singularity if we complexify the
proper-time separation at spatial infinity, so that the future is entirely
predictable from the past. If measurements at
$\mathcal{I}^{+}$ are insufficient to determine the state at
$\mathcal{I}^{-},$ one needs boundary conditions -- or another set of
`coordinates'-- on the Cauchy horizon which forms after the black hole
disappears. Boundary conditions along the
Cauchy horizon represent constraints on the initial conditions as
configurations on a final Cauchy hypersurface must be sufficient to
determine the state on an initial Cauchy hypersurface. That there is
no singularity or edge to space-time is a necessity if the path
integral for quantum gravity is to be a no-boundary wave function~\cite{HH1}.

 In the presence of a time-independent Schwarzschild black hole, say,
one commonly `Wick' rotates the real time $t$ $90^{o}$ into the lower
half-plane ($t=-i\tau$)~\cite{hartle}.
\begin{figure}[htb]
\begin{center}
\epsfig{file=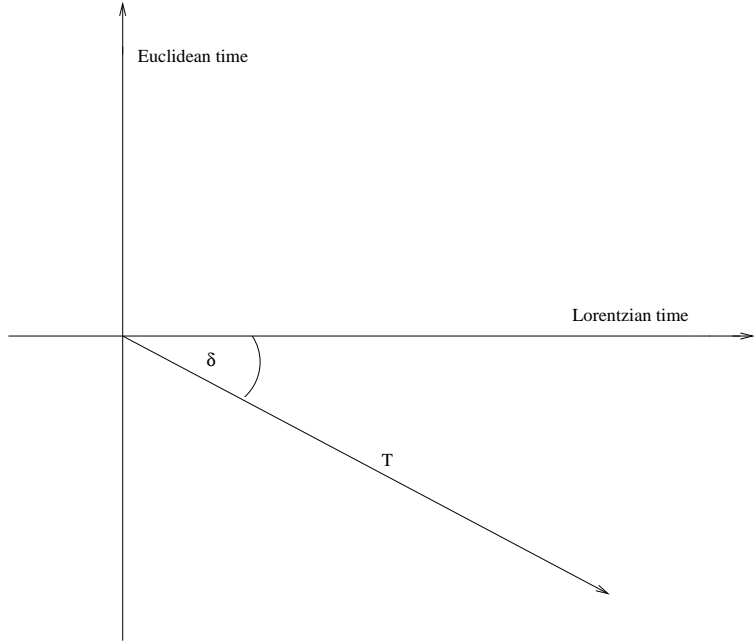,width=10cm}
\end{center}
\caption{Infinitesimal deformation of asymptotic proper-time
separation by an angle $\delta$ into the lower complex plane.\label{complex3}}
\end{figure}
 One now has a Euclidean time separation at spatial infinity and the
 space-time boundary is real and positive-definite. Where there is a future event horizon, one has the additional boundary component at
 the axis $r=2M_{0}$. In the Euclidean {\it r\'{e}gime}, a conical singularity
is present at $r=2M_{0}$ unless the Euclidean time is periodically
identified with period $8\pi M_{0}$~\cite{gibbhaw}. Thus, the coordinate
singularity at $r=2M_{0}$ as well as the curvature singularity at
$r=0$ is avoided, since $r>2M_{0}$ for Euclidean signature. The
manifold for which $r\geq 2M_{0}$ and $0\leq\tau\leq 8\pi M_{0}$ is
the real Euclidean section of the Schwarzschild solution. Boundary
conditions for physical fields then need not be specified at the
curvature singularity, which does not lie on the Euclidean section,
and these fields are also regular at the horizon. This was interpreted
as quantum cosmic censorship, and with the regularity at $r=2M_{0}$, one
 is seemingly summing over all possible configurations which lie
 inside the black hole $r<2M_{0}$. 

Hence, let us write (see fig.\ref{complex3}) 
\begin{equation}
T=|T|e^{-i\delta},\,\,0\leq\delta <\frac{\pi}{2}.\label{q38}
\end{equation}
Below we shall consider the case of infinitesimal $\delta$. Then, from
eqns.(\ref{q37}) and (\ref{q38})
\begin{eqnarray}
\Phi[\{A_{j}\};T] & = & \Phi[\{A_{j}\};|T|,\delta]\nonumber\\
& = & \hat{N}e^{-\frac{1}{2}iM_{I}|T|\cos\delta}e^{-\frac{1}{2}M_{I}|T|\sin\delta}\prod_{j}\exp\left[-\frac{1}{2}\Delta
k_{j} k_{j}|A_{j}|^{2}\coth(k_{j}|T|\sin\delta -i\varphi_{j}(|T|,\delta))\right]\nonumber\\
& = & \hat{N}e^{-\frac{1}{2}iM_{I}|T|\cos\delta}e^{-\frac{1}{2}M_{I}|T|\sin\delta}\prod_{j}\exp
(-\frac{1}{2}\left[\Omega_{Rj}+i\Omega_{Ij}\right]\Delta
k_{j} k_{j}|A_{j}|^{2}),\label{q39}
\end{eqnarray}
where 
\begin{equation}
\varphi_{j}(|T|,\delta)=-k_{j}|T|\cos\delta,\label{q40}
\end{equation}
and
\begin{equation}
\Omega_{Rj}(|T|,\delta)=\frac{\sinh (2k_{j}|T|\sin\delta
)}{2[\cosh^{2}(k_{j}|T|\sin\delta )-\cos^{2}\varphi_{j}]},\,\,\,\Omega_{Ij}(|T|,\delta)=-\frac{\sin 2\varphi_{j}}{2[\cosh^{2}(k_{j}|T|\sin\delta )-\cos^{2}\varphi_{j}]}.\label{q40'}
\end{equation}
One can also write eqn.(\ref{q39}) as
\begin{equation}
\Phi[\{A_{j}\};|T|,\delta]=\hat{N}e^{-\frac{1}{2}iM_{I}|T|\cos\delta}e^{-\frac{1}{2}M_{I}|T|\sin\delta}\prod_{j}\exp\left[-\frac{1}{2}\Delta
k_{j} k_{j}\left(\frac{1+e^{2i\varphi_{j}}\tanh r_{j}}{1-e^{2i\varphi_{j}}\tanh r_{j}}\right)|A_{j}|^{2}\right],\label{q41}
\end{equation}
where we have formally set
\begin{equation}
\tanh r_{j}(|T|,\delta)=e^{-2k_{j}|T|\sin\delta} .\label{q42}
\end{equation}
Therefore,
\begin{equation}
e^{-2r_{j}}=\tanh (k_{j}|T|\sin\delta) .\label{q43}
\end{equation}
We recognise eqn.(\ref{q41}) as the coordinate-space representation of
a quantum-mechanical squeezed
state~\cite{Matacz:tp}~\cite{Kiefer:2001wn}, with $ r_{j}(|T|,\delta)$
the squeeze parameter and $\varphi_{j}(|T|,\delta)$ the squeeze
angle. The evolution of the squeezed state is taken into account by
the $|T|$ dependence in $r_{j}$ and $\varphi_{j}$, which are in general complicated functions
of time.  Equation (\ref{q40}) is more familiar in the
limit of infinitesimal $\delta$, whence neglecting $O(\delta^{2})$
terms, $\varphi_{j}(|T|,\delta\ll
\frac{\pi}{2})\simeq-k_{j}|T|,$ corresponding to free evolution. The
high-frequency limit $k_{j}|T|\gg 1$ corresponds to $r_{j}\to 0$ for $\sin\delta\neq 0.$ 

Computing the probability density $|\Phi[\{A_{j}\};|T|,\delta]|^{2}$,
we find for small $\delta$
\begin{equation}
|\Phi[\{A_{j}\};|T|,\delta]|^{2}=|\hat{N}|^{2}e^{-M_{I}|T|\delta}\prod_{j}\exp\left[-\frac{\coth\varepsilon_{j}}{1+\frac{\sin^{2}(k_{j}|T|)}{\sinh^{2}\varepsilon_{j}}}\Delta k_{j} k_{j}|A_{j}|^{2}\right],\label{q44}
\end{equation}
where $0<\varepsilon_{j}\equiv k_{j}|T|\delta\ll 1$, i.e., $0\leq\delta\ll
 (\tilde{k}_{j}\tilde{T})^{-1}$, with $\tilde{k}_{j}=2M_{I}k_{j}$,
$\tilde{T}=\frac{|T|}{2M_{I}}$. Then, from eqn.(\ref{q43})
\begin{equation}
\varepsilon_{j}\simeq e^{-2r_{j}},\,\,\varepsilon_{j}\ll 1,\label{q45}
\end{equation}
corresponding to $r_{j}\gg 1$, which is the high-squeezing limit. 

In the squeezed-state formalism, this limit is deemed to be the classical
limit when the average number of particles in the final state is
large: $<N_{j}>=\sinh^{2}r_{j}\simeq\frac{1}{4}e^{2r_{j}}$ for
$r_{j}\gg 1$. Here, the limit of infinitesimal
$\delta$ is also a quantum-to-classical transition, from a Euclidean
($\delta\sim\frac{\pi}{2}$) theory, where there is no concept of time, to
a semi-classical Lorentzian theory, with infinitesimal $\delta$, where
the notion of a classical time parameter appears. Such issues are related
to the quantum cosmology program of Hartle and Hawking~\cite{HH1} in
the emergence of the classical universe from a smooth Euclidean
(quantum) origin, where there is no cosmological-singularity.

Indeed, the semi-classicality of the state eqn.(\ref{q37}) is
intimately related to the high-squeezing limit~\cite{Albrecht:1992kf}. If we
just consider $M_{I}$ as a fixed parameter in the theory and not a
functional of the final field configurations $\{x_{j}\}$, then the WKB
condition is met when 
\begin{equation} 
\left|\frac{\Omega_{Ij}}{\Omega_{Rj}}\right|=\left|\frac{\sin2\varphi_{j}}{\sinh(2k_{j}|T|\sin\delta
)}\right|=|\sin2\varphi_{j}\sinh 2r_{j}|\gg 1,\label{q45'}
\end{equation}
which is satisfied in the high-squeezing limit $\varepsilon_{j}\ll 1$
even if $\sin2\varphi_{j}=0.$ The final state of the remnant Hawking flux, therefore, becomes more
classical in the WKB sense in the limit $\delta\to 0$. In this limit,
one can effectively consider the final perturbations as being
represented by a classical probability distribution
function~\cite{grishchuk}~\cite{Albrecht:1992kf}~\cite{Polarski:1995jg}.
As in the inflationary scenario, the perturbations on the black-hole background space-time, which had a
quantum-mechanical origin, cannot be distinguished from classical
stochastic perturbations, without the need of an environment for
decoherence. It is the field amplitudes which one observes, and these
amplitudes have a classical phase-space distribution. The squeezed
nature of the final quantum state for the stochastic
gravitational-wave background produced by inflation is not stationary~\cite{grishchuk}.

The initial conditions for the perturbations in the black-hole case
also have an inflationary analogue. In cosmology, the assumption is
that at some early `time' just prior to inflation, the modes are in
their adiabatic ground state. This originates from the assumption that
the universe was in a maximally-symmetric state at some time in the
past~\cite{HH1}. A similar assumption obtains in our black-hole case,
where we assumed that the initial
perturbations were very weak so that the initial matter and its
gravitational field were spherically symmetric.  

 We assume that $\tilde{T}\gg 1$; that is, we observe the black
hole at infinity at times much greater than its collapse timescale,
which is of order $\pi M_{I}$~\cite{MTW}. If $\tilde{k}_{j}$ is
moderately large, then for small $\varepsilon_{j}$ (keeping
$O(\varepsilon_{j}^{2})$ terms)
\begin{eqnarray}
|\Phi[\{A_{j}\};|T|,\delta]|^{2} & = & |\hat{N}|^{2}e^{-M_{I}|T|\delta}\prod_{j}\exp\left[-\frac{\varepsilon_{j}}{\varepsilon^{2}_{j}+\sin^{2}(k_{j}|T|)}\Delta k_{j} k_{j}|A_{j}|^{2}\right]\nonumber\\
& \stackrel{=}{\varepsilon_{j}\to 0} & |\hat{N}|^{2}\prod_{j}\exp\left[-\Delta k_{j} k_{j}|A_{j}|^{2}\sum_{n=1}^{\infty}\Delta\omega_{n}\delta (k_{j}-\omega_{n})\right],
\label{q46}
\end{eqnarray} 
where $\omega_{n}=\frac{n\pi}{|T|}$, $\Delta\omega_{n}=\frac{\pi}{|T|}$ and
we used $k_{j}>0$ with $k_{j}|A_{j}|^{2}\to 0$ as $k_{j}\to 0$ to
convert the sum $\sum_{n=-\infty}^{\infty}$ into
$\sum_{n=1}^{\infty}$. Hence, assuming the interchange of $j$ and $n$
sums in the continuum limit for the $\{k_{j}\}$ frequencies is valid, then 
\begin{equation}
|\Phi[\{A_{j}\};|T|,\delta\to 0]|^{2}=|\hat{N}|^{2}\prod_{slmP}\prod_{n=1}^{\infty}e^{-\Delta\omega_{n}\omega_{n}|A_{snlmP}|^{2}},\label{q47}
\end{equation}
 so the frequencies $\{\omega_{n}\}$ dominate
the final state. This is the same as in ref.~\cite{me} where we
used contour integration to obtain the probability density. Equation
(\ref{q46}) describes a Gaussian non-stationary process in that the
variance is an oscillatory function of the asymptotic proper time. Rather than travelling waves, one is now dealing with
standing bosonic waves, where the amplitudes for left- and right-moving waves are large and almost equal, similar to the cosmological
scenario~\cite{grishchuk} -- see eqn.(\ref{q1'}). These standing waves imply a correlation
between particles with opposite frequencies (and azimuthal angular
momentum $m$) in the final state, indicative of a pure
state. We suggest that the discrete frequencies
$\{\omega_{n}\}$ are, in principle, observable feature of the relic
Hawking radiation, either directly or indirectly. 

The presence of the delta function in eqn.(\ref{q46}) indicates that
in the high-squeezing limit, the random variable $\varphi_{j}$
associated with the final state is squeezed to discrete values
independently of the quantum numbers $\{slmP\}$. Note that it is only
the squeeze phases $\{\varphi_{j}\}$ of the perturbations which are
fixed and correlated in the high-squeezing limit. There are phases in
our theory, however, which we deem to have a random and even
distribution; namely, the phases associated with the initial state of
the fluctuations incoming from ${\mathcal{I}^{-}}$, and the phases
associated with the spatial field distribution labelled by $\{a_{sklmP}\}$. The final field amplitudes for the remnant Hawking radiation reaching the null
hypersurface ${\mathcal{I}^{+}}$ $\{b_{j}\}$ can be matched on
$\Sigma_{F}$ with the amplitudes $\{a_{j}\}$ associated with the
space-like hypersurface $\Sigma_{F}$~\cite{me}. The $\{b_{j}\}$ are in turn connected, through a
Bogoliubov transformation, with amplitudes $\{c_{j'}\}$ for
travelling waves incoming from ${\mathcal{I}^{-}}$. A
positive-definite probability distribution can be obtained by averaging over the random phase of the $\{c_{j'}\}$
fluctuations. Therefore, it is only the phases $\{\varphi_{j}\}$
associated with the standing waves which do not have a random
character.

The normalisation factor $|\hat{N}|^{2}$ is determined by integrating
over the dimensionless variables $\{y_{j}\}$, where
$x_{j}=\Delta k_{j} k_{j}|A_{j}|^{2}\equiv M_{I}^{2}y_{j}$, so that
the sum of all probabilities of all possible configurations $\{y_{j}\}$ is
unity. Treating the initial ADM mass $M_{I}$ as a fixed parameter rather
than as a functional
of $\{x_{j}\}$, we find that
\begin{eqnarray}
|\hat{N}|^{2} & = & \prod_{j}\frac{M_{I}^{2}\coth\varepsilon_{j}}{1+\frac{\sin^{2}(k_{j}|T|)}{\sinh^{2}\varepsilon_{j}}}\nonumber\\
& = & \prod_{j}[\cosh 2r_{j}-\cos 2\varphi_{j}\sinh 2r_{j}]^{-1}M_{I}^{2}.\label{q48}
\end{eqnarray}
In fact, in the high-squeezing limit $\delta\to 0$, final states which
differ by a coordinate-dependent phase $e^{-\frac{1}{2}iM_{I}T}$ are
deemed to be physically equivalent. If we considered the ADM mass as a
functional of $\{x_{j}\}$, then the normalisation factor would be
altered for finite $\delta$, as would the value of the entropy -- see below. Thus, there is an
ambiguity in the normalisation factor. In addition, this is related to the
ambiguity in the form of the entropy in the high-squeezing limit with
and without a surface term included in the Lagrangian~\cite{Matacz:tp}. In the limit $\varepsilon_{j}\to 0$, then
\begin{equation}
|\Phi[\{A_{j}\};|T|,\delta\to 0]|^{2}=\left[\prod_{j}M_{I}^{2}\sum_{n=1}^{\infty}\Delta\omega_{n}\delta (\omega_{n}-k_{j})\right]\prod_{slmP}\prod_{n=1}^{\infty}e^{-\Delta\omega_{n}\omega_{n}|A_{snlmP}|^{2}}.\label{q48'}
\end{equation}
This equation is reminiscent of the Wigner function corresponding to
the vacuum state in the limit of infinite squeezing~\cite{Polarski:1995jg},
giving the classical `trajectories' of the system. The Wigner function
cannot in general be interpreted as a classical probability density for finite
squeezing, that is, finite $\delta$, except for Gaussian states. 

A consequence of the high-squeezing behaviour is that the variance in
the amplitudes $\{x_{j}\}$ is large, so that there are large
statistical deviations of the observable power spectrum from its
expected value. This is just a manifestation of the uncertainty
principle. Indeed, with respect to the first expression in eqn.(\ref{q46}), and
using eqn.(\ref{q48}), we find that
\begin{eqnarray}
<y_{j}>_{|\Phi |^{2}} & \stackrel{=}{\varepsilon_{j}\to 0} & \left[M_{I}^{2}\sum_{n=1}^{\infty}\Delta\omega_{n}\delta (\omega_{n}-k_{j})\right]^{-1}\nonumber\\
& \equiv & [\rho_{j}(M_{I})]^{-1},\label{q48''}
\end{eqnarray}
where $<>_{|\Phi |^{2}}$ denotes the expectation value with respect to
the probability density $|\Phi |^{2}.$ It is reasonable to suggest that $\rho_{j}(M_{I})$ represents a microcanonical density-of-states.

In inflationary cosmology, the oscillation phases of standing waves
have fixed values, giving rise to zeros in the power spectrum, which are
characteristic of the CMBR. In this case, the power spectrum of the
cosmological perturbations in the present universe is not a smooth
function of frequency. The standing-wave pattern, due to squeezing,
induces oscillations in the power spectrum. This in turn produces
Sakharov oscillations~\cite{Albrecht:1992kf}\cite{Albrecht:1996dq}, produced by
metric and scalar perturbations, in the distribution of higher-order multipoles
$l(l+1)C_{l}$ of the angular correlation function for the temperature
anisotropies~\cite{Grishchuk:1995ia}~\cite{Bose:2001fa} in the CMBR
for all perturbations at a fixed time whose wavelength is of the order of or greater than the
Hubble radius defined at this time. That is, the peaks and troughs of
the angular power spectrum have a close relationship with the maxima
and zeros of the metric power spectrum. However, for long wavelengths, the
power spectrum is sufficiently smooth. Sakharov oscillations,
therefore, exist as a result of the squeezed nature of the scalar and
metric perturbations.  

\section{Entropy and Squeezing}\label{classical}

There have been many accounts of how to determine the entropy
generation in the squeezing formalism~\cite{Gasperini:1993mq}~--~\cite{Matacz:tp}. Hu and Pavon~\cite{pavon} were the first to associate entropy
generation with the monotonic increase in the average particle number
with time, induced by parametric amplification in a vacuum
cosmological space-time. As squeezing is the quantum analogue of
parametric amplification, one would expect that the squeezed-state
formalism can compute entropy production. This is indeed the case,
although, as with any entropy calculation, the nature of the
coarse-graining must be specified. For squeezing, this is particularly
relevant as squeezed evolution is unitary, i.e., there is no loss of
information in principle in the evolution of the initial pure state to
the final pure squeezed quantum state. How one chooses to measure the
observables associated with the final squeezed state determines the
entropy. One can reduce the final density matrix with respect to a
Fock or coherent state basis~\cite{Matacz:tp}, or use eigenstates of
the superfluctuant variable~\cite{Gasperini:1993mq}~\cite{Gasperini:1992xv}~\cite{Gasperini:1995yd}.
In refs.~\cite{Gasperini:1993mq}~\cite{Gasperini:1992xv}~\cite{Gasperini:1995yd},
the loss of information comes from the increased dispersion of the superfluctuant operator.

In the classical limit of large average particle number, corresponding
to the large-squeezing {\it r\'{e}gime} $r_{j}\gg 1$, a universal form
for the entropy density growth $\Delta S_{j}$ obtains for each mode~\cite{Gasperini:1993mq}~--~\cite{Matacz:tp}:
\begin{equation}
\Delta S_{j}\simeq 2r_{j},\,\,r_{j}\gg 1,\label{qa}
\end{equation} 
irrespective of the particular coarse-graining. In particular, on
averaging over the squeezing angle $\varphi_{j}$ of each Fourier mode,
one obtains
eqn.(\ref{qa})~\cite{Gasperini:1993mq}~\cite{Prokopec:1992ia}~\cite{Brandenberger:1992sr}~--~\cite{Kruczenski:1994pu}.
One can determine whether coarse-graining with respect
to the squeeze angle is appropriate from the prescription discussed in
ref.~\cite{Kiefer:2001wn}. Calculating the
entropy $S$ from eqn.(\ref{q44}) using eqn.(\ref{q48}), then
\begin{eqnarray}
S & = & -\int\prod_{j}dy_{j}P(y_{j})\ln P(y_{j})\nonumber\\
& = & 1+\sum_{j}\ln\left[M_{I}^{-2}(e^{2r_{j}}\sin^{2}\varphi_{j}+e^{-2r_{j}}\cos^{2}\varphi_{j})\right].\label{qb}
\end{eqnarray} 
Thus, the entropy eqn.(\ref{qb}) comes from our ignorance of the final
radiation configuration. The constant term is not important. In the high-squeezing limit
\begin{equation}
S\simeq 1-\sum_{j}\ln M_{I}^{2}+2\sum_{j}r_{j}+\sum_{j}\ln\sin^{2}\varphi_{j}.\label{qc}
\end{equation} 
Evidently, $\sin\varphi_{j}=0$ for particular values of the
frequency. Even if $\sin\varphi_{j}\to 0$, for $r_{j}\to\infty$ and
ignoring constant terms, then one may argue that $\Delta S_{j}\simeq 2r_{j}$. However, an alternative derivation keeping
$O(\varepsilon_{j}^{2})$ terms gives from eqn.(\ref{q48})
\begin{eqnarray}
S & = & 1-\sum_{j}\ln[M_{I}^{2}\sum_{n=1}^{\infty}\Delta\omega_{n}\delta
(\omega_{n}-k_{j})]\nonumber\\
& = & 1-\sum_{j}\ln\rho_{j}(M_{I}).\label{qc'}
\end{eqnarray} 

How we take the high-squeezing limit is therefore important. The
effect of the final term in eqn.(\ref{qc}) is to reduce the entropy
from the maximal value given in eqn.(\ref{qa}). In the cosmological
scenario, the primordial gravitational-wave background entropy is
significantly smaller than eqn.(\ref{qa}) due to the information about
the initial inflationary state of the universe manifesting itself in
the primordial peaks in the multipole spectra of the CMB temperature
anisotropy~\cite{Kiefer:1999sj}. The presence of these peaks, therefore,
is incompatible with a totally random squeezing angle, and so a
coarse-graining with respect to $\varphi_{j}$ is not possible. A
similar conclusion may be argued in our case. We will comment on this
in another paper~\cite{me}, where we discuss the discreteness of the
radiation frequencies in the context of Bekenstein's discrete
event-horizon-area theory, and how information about the initial state
may be contained in the spectral lines.   

\subsection{Classical Predictions}

We now discuss how strong peaks in the wave function lead to definite
predictions. In quantum cosmology, wave functions are commonly
peaked about correlations between coordinates and momenta. In
ref.~\cite{halliwell}, identification of such correlations came {\it
via} the Wigner function. If the Wigner function
$W(p,q)$ factorised into a function of position $q$ and a function of momentum $p$, then no
correlation between $q$ and $p$ is predicted. When $W(p,q)$ is peaked
about some region in phase space $p=f(q)$, then the wave function
predicts this particular correlation.

An alternative proposal for measuring correlations was given
in ref.~\cite{anderson}. Here, projection onto coherent states, where
momentum and position are equally known, as in classical theory, is
employed for predicting classical correlations from a general Wigner
function. It was shown that in the harmonic-oscillator case, the correlation
between $p$ and $q$ was such that the Hamiltonian equalled the
classical energy. We arrive at similar conclusions. In
ref.~\cite{halliwell}, the predicted correlation was between the
Hamiltonian equated to the quantum energy. The problem with this
approach was that crude approximate Wigner functions bore no resemblance to the exact Wigner
functions. How Wigner functions appear in the context of our
two-surface formalism will be considered in another paper.

If the quantum state $\Psi$ is `sufficiently' peaked about a region in
the phase space, we observe correlations between the observables which
characterise this region. Phase-space configurations for which
$\Psi$ is small are precluded and will not be observed. Where $\Psi$
is neither small nor sufficiently peaked, no predictions can be
made~\cite{habib1}. These conclusions are drawn from a new interpretation of
quantum mechanics where wave functions are not associated with
probabilities.

With the ADM mass $M_{I}$ as a parameter, we look for predictions from the Heisenberg picture wave
functional $\Psi^{(H)}_{n_{j}}(x_{j})=\frac{N}{\pi}e^{-\frac{1}{2}x_{j}}L_{n_{j}}(x_{j}).$ Were we to restore our units, then the argument of the
Laguerre polynomial would be proportional to $\hbar^{-1}$. Thus, in
the limit of large argument, we use for large $x$~\cite{grad}
\begin{equation}
L_{k}(x)\simeq \frac{(-x)^{k}}{k!}.\label{q28}
\end{equation}   
From a dimensional argument, $x_{j}\propto\frac{M_{I}^{2}}{m_{pl}^{2}}\gg
1$. In this case also,
therefore, the approximation in eqn.(\ref{q28}) applies. One can now find the
peak in the wave function as a function of $x_{j}$ at 
\begin{equation}
n_{j}=\frac{1}{2}k_{j}\Delta k_{j}|A_{j}|^{2}\hbar^{-1}.\label{q29}
\end{equation} 
Considering the spin-0 case, for example, from eqn.(\ref{q7}) and
eqn.(\ref{q29})
\begin{equation}
n_{j}=2\pi k_{j}\Delta k_{j}|z_{j}|^{2}|a_{j}+(-1)^{s}Pa_{s,-k_{j} lmP}|^{2}\label{q30}.
\end{equation}  
In a previous paper~\cite{me}, we showed that for spin-0 perturbations
\begin{equation}
|b_{j}|^{2}=2\pi k_{j}|z_{j}|^{2}|a_{j}+(-1)^{s}Pa_{s,-k_{j} lmP}|^{2}\label{q31},
\end{equation}  
where the $\{b_{j}\}$ are Fourier amplitudes associated with the
radiation reaching ${\mathcal{I}^{+}}$; i.e., eqn.(\ref{q31}) matches
the positive-frequency decomposition for massless spin-0 particles reaching
${\mathcal{I}^{+}}$ (travelling waves) with our two-surface method for field
configurations on the $\{t=T\}$ hypersurface $\Sigma_{F}$ (standing
waves). Hence, $n_{j}=\Delta k_{j}|b_{j}|^{2}.$ Thus we find that
\begin{equation}
\sum_{j}\hbar n_{j}k_{j}=\sum_{j}\Delta k_{j}k_{j}|b_{j}|^{2}=M_{I}.\label{q32}
\end{equation}  
The right-hand side is just the final total energy in the massless
spin-0 fluctuations, which equals the initial ADM mass $M_{I}$, as
demanded by our boundary-value problem. The left-hand
side is just the total energy of the radiated particles.

\section{Conclusion}

In this paper, we have illustrated many aspects of the two-surface
formulation for linearised integer-spin fields propagating in an
evaporating black-hole space-time. When the proper-time separation $T$
at infinity between the initial and final hypersurfaces is deformed
infinitesimally into the lower complex plane, and one has
spherically-symmetric initial fields, one obtains a quantum-mechanical
squeezed-state formalism. The large squeezing limit is equivalent to
the WKB limit and corresponds to an infinitesimal angle in the lower
complex-$T$ plane. This we believe is related to the emergence of time
in a semi-classical Lorentzian space-time from a timeless Euclidean
{\it r\'{e}gime}. As the highly-squeezed final state is a pure state,
our complexification technique has seemingly avoided the
unpredictability associated with the momentarily-naked singularity,
which is believed to be present prior to the complete disappearance of
the black hole.
 
As in the cosmological scenario, we found that the bosonic
perturbations on the black-hole background can be deemed to be a
stochastic collection of standing waves rather than travelling waves
in the high-squeezing limit. This leads to the prediction of peaks in
the power spectrum of the relic Hawking radiation analogous to the Sakharov
oscillations in the CMBR. 

Many of the features discussed in this paper may be valid in the
instance of non-spherical initial fields; i.e., non-zero initial
perturbations. In this case, there will be cross terms in the classical action involving the initial and final
field configurations. One can subsequently diagonalise the classical action
resulting in a product of squeezed states, and so the results in this
paper are valid. For rotating black holes, similar conclusions
to those arrived at in this paper are possible. One must also include
the fermionic perturbations, which are described by anti-commuting Grassmann variables.

\section*{Acknowledgements}

I thank Peter D'Eath and Ian Moss for helpful discussions, and the EPSRC for funding.
 
\appendix
\section*{Appendix A -- Derivation of the wave function eqn.(\ref{q6})}\label{A}

\indent

To arrive at eqn.(\ref{q6}), we set $a=e^{-ik_{j}T}$, $y=0$ and  $x=
x_{j}$ in Mehler's formula~\cite{erdelyi}
\begin{equation}
(1-a^{2})^{\frac{1}{2}}\sum_{p=0}^{\infty}e^{-\frac{1}{2}(x^{2}+y^{2})}\frac{a^{p}H_{p}(x)H_{p}(y)}{2^{p}p!}=e^{\frac{1}{2}(x^{2}-y^{2})}e^{-\frac{(x-ay)^{2}}{1-a^{2}}},\label{c2}
\end{equation}
where 
\begin{equation}
H_{p}(x)=(-1)^{p}e^{x^{2}}\frac{d^{p}}{dx^{p}}e^{-x^{2}}\label{c3}
\end{equation}
 are Hermite polynomials which satisfy 
\begin{equation}
\frac{1}{2^{p}p!\pi^{\frac{1}{2}}}\int_{-\infty}^{\infty}dx\,e^{-x^{2}}[H_{p}(x)]^{2}=1.\label{c4}
\end{equation}
 Then,
\begin{equation}
[2\pi i\sin(k_{j}T)]^{-\frac{1}{2}}e^{\frac{1}{2}ix_{j}^{2}\cot(k_{j}T)}=\sum_{p=0}^{\infty}e^{-iE_{p}T}\psi_{p}(x_{j})\psi_{p}(0),\label{sum}
\end{equation}
where $E_{p}=k_{j}(p+\frac{1}{2})$ and
\begin{equation}
\psi_{p}(x_{j})=\frac{e^{-\frac{1}{2}x_{j}^{2}}H_{p}(x_{j})}{(2^{p}p!\pi^{\frac{1}{2}})^{\frac{1}{2}}},
\end{equation}
i.e. $$\int_{-\infty}^{\infty}dx_{j}|\psi_{p}(x_{j})|^{2}=1.$$
Only even terms contribute to the sum in eqn.(\ref{sum}) as~\cite{grad}  
\begin{eqnarray}
H_{2p}(0) & = & \frac{(-1)^{p}(2p)!}{p!},\label{c9} \\
H_{2p+1}(0) & = & 0\label{c10}.
\end{eqnarray}
 Hence, 
\begin{equation}
[2\pi i\sin(k_{j}T)]^{-\frac{1}{2}}e^{\frac{1}{2}ix_{j}^{2}\cot(k_{j}T)}=\sum_{p=0}^{\infty}e^{-iE_{2p}T}\psi_{2p}(x_{j})\psi_{2p}(0).
\end{equation}
In addition, set $x_{j}= (\frac{k_{j}}{V})^{\frac{1}{2}}Re(A_{j}),$
$y_{j}= (\frac{k_{j}}{V})^{\frac{1}{2}}Im(A_{j}).$ Then,
\begin{eqnarray}
\frac{e^{\frac{1}{2}ik_{j}V^{-1}|A_{j}|^{2}\cot(k_{j}T)}}{2\pi
i\sin(k_{j}T)}\! & = &\!
\sum_{p,p'=0}^{\infty}e^{-i(E_{2p}+E_{2p'})T}\psi_{2p}(x_{j})\psi_{2p'}(y_{j})\psi_{2p}(0)\psi_{2p'}(0)
\nonumber \\
\! & = & \!\sum_{p=0}^{\infty}e^{-2iE_{p}T}\!\sum_{p'=0}^{p}\psi_{2p'}(x_{j})\psi_{2p-2p'}(y_{j})\psi_{2p'}(0)\psi_{2p-2p'}(0)
\nonumber \\ 
\! & = & \!\frac{e^{-\frac{k_{j}}{2V}|A_{j}|^{2}}}{\pi}\!\sum_{p=0}^{\infty}\frac{(-1)^{p}e^{-2iE_{p}T}}{2^{2p}p!}\sum_{p'=0}^{p}\frac{p!}{p'!(p-p')!}H_{2p'}(x_{j})H_{2p-2p'}(y_{j}),\nonumber
\end{eqnarray}
where $V^{-1}=\Delta k_{j}$. In the limit $|T|\rightarrow 0_{+}$ (or rather $k_{j}|T|\ll 1$), as the eigenfunctions $\psi_{p}(x_{j})$ form a complete orthonormal set 
\begin{eqnarray}
\lim_{k_{j}|T|\rightarrow 0_{+}}[2\pi
i\sin(k_{j}T)]^{-1}e^{i\frac{k_{j}}{2V}|A_{j}|^{2}\cot(k_{j}T)} & = &
\delta(x_{j})\delta(y_{j}) \nonumber \\
& \equiv & \delta^{(2)}((\frac{k_{j}}{V})^{\frac{1}{2}}A_{j}).\label{delta}
\end{eqnarray}
This suggests that $A_{j}\to 0$ as $k_{j}|T|\rightarrow 0_{+}$,
agreeing with our initial conditions of very weak perturbations. We now use the identity eqn.(\ref{q12}). Taking the product
over all $j$ and introducing a normalisation factor $\hat{N}$, then
\begin{eqnarray}
\Psi[\{a_{j}\};T] & = & \prod_{j}\Psi_{j}(a_{j};T)\nonumber\\
& = & \hat{N}e^{-i\frac{1}{2}M_{I}T}e^{-\frac{1}{2}\sum_{j}\frac{k_{j}}{V}|A_{j}|^{2}}\prod_{j}\sum_{p=0}^{\infty}e^{-2iE_{p}T}L_{p}\left(\frac{k_{j}}{V}|A_{j}|^{2}\right),\label{fho1}
\end{eqnarray}
which is eqn.(\ref{q6}), where we have included the contribution from
the time-like boundary. One can confirm that eqn.(\ref{fho1}) gives us the
first part of eqn.(\ref{q6}) by simply using the generating function for
Laguerre polynomials~\cite{grad}.

\end{document}